\documentclass[12pt]{article}
\usepackage{graphicx}
\usepackage{amssymb}
\usepackage{amsmath}
\begin{document}

\title{\bf{Classical Simulation of Quantum Fields II}}

\author{T. Hirayama$^a$, B. Holdom$^a$, R. Koniuk$^b$, T. Yavin$^b$
\vspace*{2ex}\\
\emph{\small  ${}^a$ Department of Physics, University of Toronto}\\[-1ex]
\emph{\small  Toronto ON Canada M5S1A7}\\
\emph{\small  ${}^b$ Department of Physics and Astronomy, York University}\\[-1ex]
\emph{\small  Toronto ON Canada M3J1P3}}
\date{}
\maketitle
\begin{abstract}
We consider the classical time evolution of a real scalar field in 2 dimensional
Minkowski space with a $\lambda \phi^4$ interaction. We compute the spatial and temporal two-point correlation functions and extract the renormalized mass of the interacting theory.
We find our results are consistent with the one- and two-loop quantum computation.  We also perform Monte Carlo simulations of the quantum theory and conclude that the classical scheme is able to produce more accurate results with a fraction of the CPU time.
\end{abstract}

\section{Introduction}

In quantum field theory, the vacuum is filled with zero-point energy to which each mode contributes $\hbar\omega/2$. This picture raises the question, what dynamics would emerge in an interacting classical field theory if each mode is initialized with energy $\hbar\omega/2$. The dynamics of modes scattering and re-scattering off one another resembles the corresponding picture in quantum field theory. If we assume the phase of each mode is randomly distributed, then Lorentz invariant correlation functions arise upon averaging over phases. For example if we consider $\lambda\phi^4$ theory, the value of $\phi^2$ after averaging over phases equals the vacuum expectation value $\langle 0|\phi^2(x)|0\rangle$ of quantum field theory. This vacuum expectation value in combination with the $\lambda\phi^4$ term produces a mass correction that is exactly the one-loop correction in quantum field theory.

Classical zero-point fluctuations of \textit{free massless} scalar and vector fields, the Lorentz invariance thereof, and the effects of phase averaging, have been considered before, most notably by Boyer \cite{Boy89}. By studying correlation functions of random electromagnetic fields, and the behavior of charged systems in their presence, Boyer has emphasized interesting similarities to quantum mechanics.

In an accompanying paper, two of the present authors have considered the dynamics of background fluctuations with an $\hbar\omega/2$ energy spectrum in interacting classical field theory. They showed that a lattice formulation allows for a nonperturbative analysis of the classical evolution equations. They found critical, strong coupling behavior in $\lambda \phi^4$ theory that is strikingly similar to what is known about the quantum field theory. They also developed a perturbative expansion of the classical theory in the continuum and found a loop expansion very similar to, but apparently not identical to, that of quantum field theory. All of this motivates the present work, where we attempt a more quantitative comparison of the classical and quantum theories, where both are defined on the lattice.

In the classical theory
it is easy to implement the time evolution of the system, and we
expect the configuration at a later time to incorporate the effects of
interactions. From the 2-point correlator we can obtain the physical mass, to be compared with the physical mass obtained by conventional methods from the lattice quantum theory. Our interest here is not in the continuum limit of these lattice theories, or the corresponding critical behavior. We wish to compare physical masses that are safely above the inverse size of the systems, and safely below the inverse lattice spacing. Our focus is on the loop corrections to these masses. This means that effects at two-loops or higher will need to be extracted, since we already know that the one-loop effects of the two theories agree.

There are extensive studies of classical field theories and their simulation in the literature, typically to approximate quantum field theories in thermal backgrounds or other backgrounds with high occupation number  \cite{thermdecay,therminit,thermstable}. In some of these studies one component of the initialized classical field does model the zero point spectrum of the quantum vacuum \cite{therminit}. But even then the focus has been on the dynamics of the thermal component, and not on how the interactions are manifested on the vacuum part. One of the common objects of study is the plasmon (or Landau) damping rate, related to the imaginary part of the on-shell self-energy \cite{thermdecay}. Instead we compare to zero temperature field theory and study how closely the classical theory simulates quantum mass renormalizations, and thus our interest is in the real part of the self-energy.

\section{Time Evolution}

We study a classical $\lambda\phi^4$ theory in 1+1 Minkowski space on a lattice. The action is
\begin{eqnarray}
 S &=& \sum_i \sum_{j=0}^{N-1} \frac{\dot{\phi}^2(i,j)}{2} 
  -\frac{(\phi(i,j+1)-\phi(i,j))^2}{2}
  -\frac{m_0^2}{2}\phi^2(i,j) -\frac{\lambda}{4}\phi^4(i,j),
\end{eqnarray}
where we employ the units $\hbar=c=a=1$, and $a$ is the lattice
spacing along the $x$ direction. The integers $i$ and $j$ label the
sites in the $t$ and $x$ directions, and $j$ runs from $0$ to $N-1$. A periodic boundary condition is
imposed along the $x$ direction, i.e. $\phi(i,N)=\phi(i,0)$. The equations
of motion are obtained from the action and we adopt a leapfrog method
to numerically integrate them.
\begin{eqnarray}
 \phi(i+1,j) &=& \phi(i,j)+a_t\dot{\phi}(i+1/2,j),
  \label{eq1}\\
 \dot{\phi}(i+3/2,j)&=& \dot{\phi}(i+1/2,j)
  +a_t\Big(\phi(i+1,j+1)-2\phi(i+1,j)+\phi(i+1,j-1)\nonumber\\
 &&
  -m_0^2\phi(i+1,j)-\lambda\phi^3(i+1,j)\Big)
  \label{eq2}
\end{eqnarray}
 The leapflog scheme can be shown to be reversible and
has second order accuracy.   We obtain accurate classical evolution with $a_t=0.1$, where $a_t$ is the lattice spacing along the time direction. (Decreasing $a_t$ didn't significantly change our results.)

In quantum field theory, each mode contributes $\hbar\omega/2$ to the
zero-point energy and thus the initial configuration of interest is
\begin{eqnarray}
 \phi_0(i,j) &=&\sum_{k=-N/2}^{N/2-1} \frac{1}{\sqrt{N\omega_k}}
  \cos(\omega_k a_t i+\frac{2\pi k}{N}j+\theta_k ),
  \label{ph0}
  \\
 &&\omega_k =\sqrt{4\sin^2(\pi k/N)+\mu^2},
\end{eqnarray}
where $\theta_k $ is a phase parameter, and $\omega_k $ is the lattice
dispersion relation. We leave the mass $\mu$ 
as a parameter since the physical mass will be different from the bare
mass $m_0$. One can easily check that the energy of each mode
equals $\hbar\omega_k /2$ 
(with $\lambda=0$ and $\mu=m_0$). 

The phases $\theta_k $ are independent
parameters and we assume each $\theta_k $ is randomly distributed over
$[0, 2\pi)$. Then, for example, we obtain the following average values
\begin{eqnarray}
 \langle \cos\theta_k \cos\theta_k' \rangle &\equiv&
  \int\frac{d\theta_k }{2\pi}\int\frac{d\theta_k' }{2\pi}
  \cos\theta_k \cos\theta_k' =
  \frac{1}{2}\delta_{kk'}.
\end{eqnarray}
We shall use $\langle \cdot\rangle$ to denote this average over the phases.
This averaging gives an interesting indication of the emergence of quantum
effects if we compute the average values of $\phi_0(i,j)\phi_0(i',j')$,
\begin{eqnarray}
 \langle \phi_0(i,j)\phi_0(i',j')\rangle
  &=&\sum_{k=-N/2}^{N/2-1}
  \frac{1}{{2N\omega_k }}\cos(\omega_k a_t(i-i')+\frac{2\pi k}{N}(j-j'))
  \label{free}\\
 &=&{\rm Re}\langle 0|T\phi(i,j)\phi(i',j')|0\rangle,
\end{eqnarray}
where the r.h.s of the second line is the real part of the 2-point function in
lattice quantum field theory, i.e. the real part of the Feynman propagator. 
This implies that the mass shift induced from the interaction term of the classical action is exactly the same
as the one-loop mass correction in quantum field theory.

We shall use
(\ref{ph0}) as the starting point for the investigation of the interacting theory.
The key point is that the
classical evolution of classical fields can be easily carried out in the
presence of interactions and the configuration at a later time incorporates the interacting dynamics. We can use (\ref{ph0}) to specify
the initial conditions, $\phi(0,j)=\phi_0(0,j)$ and 
$\dot{\phi}(1/2,j)=\dot{\phi}_0(1/2,j)$, but evolve forward in time with the
full $\lambda\neq 0$ equations of motion (\ref{eq1}) and (\ref{eq2}).
We then compute correlators
\begin{eqnarray}
 \langle \phi(i,j)\phi(i',j')\rangle
\end{eqnarray} 
where $i$ and $i'$ are sufficiently far away from the initial
time. We obtain the average values
by repeatedly computing 
$\phi(i,j)\phi(i',j')$ starting from different sets of random phases. To reduce
statistical noise we make use of translational invariance to also average over space.

Various correlation functions can be calculated with respect to the final time slice of the simulation, at $t_f=a_t i_f$. The time correlator $G_t(i)$, the space correlator $G_x(j)$, and the zero-mode time
correlator $G_0(i)$ are defined as follows.
\begin{eqnarray}
 G_t(i)&\equiv&\left\langle 
  \frac{1}{N}\sum_{j=0}^{N-1}
  \phi(i_f,j)\phi(i_f-i,j)\right\rangle\\
 G_x(j)&\equiv& \left \langle
  \frac{1}{N}\sum_{j'=0}^{N-1}
  \phi(i_f,j')\phi(i_f,j'+j)\right\rangle\\
 G_0(i)&\equiv&\left\langle \frac{1}{N^2} \sum_{j, j'=0}^{N-1}
  \phi(i_f,j)\phi(i_f-i,j')\right\rangle
\end{eqnarray}
Any of these correlators can be used to extract mass, by fitting to the corresponding free field quantities.
\begin{figure}
  \begin{center}
   \includegraphics[scale=0.6]{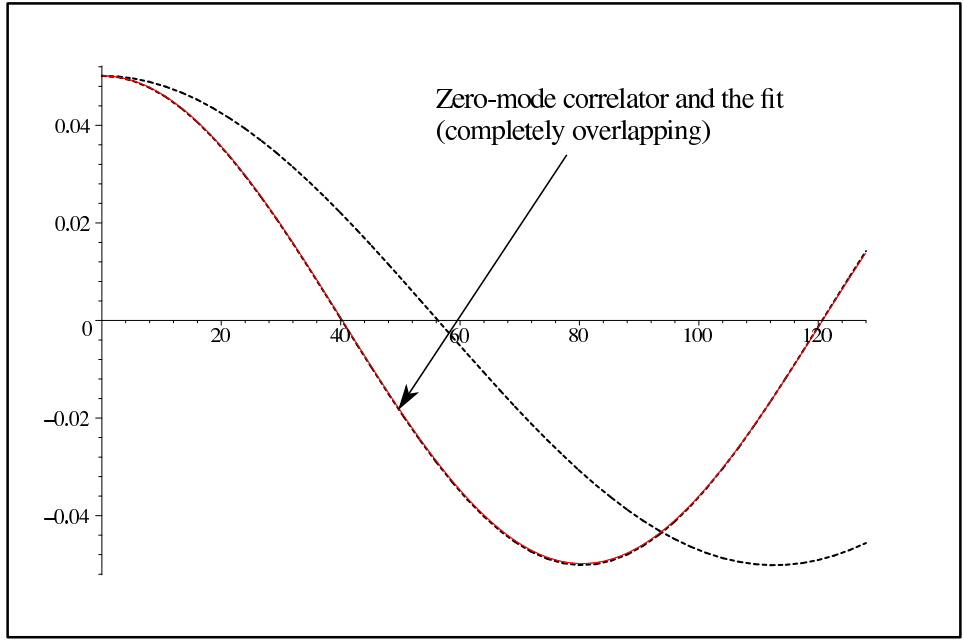}
  \end{center}
  \vspace{-3ex}
\caption{{\bf The zero-mode correlation function versus position in lattice units.} The zero-mode correlator from the classical simulation for $m_0^2=51/N^2$ and $\lambda=20/N^2$. It is well fit by analytical free correlator with $\mu^2=100/N^2$ (the lines completely overlap). Also shown is the free correlator with $\mu^2=m_0^2$.}
  \label{figa}
\end{figure}
\begin{figure}
  \begin{center}
   \includegraphics[scale=0.6]{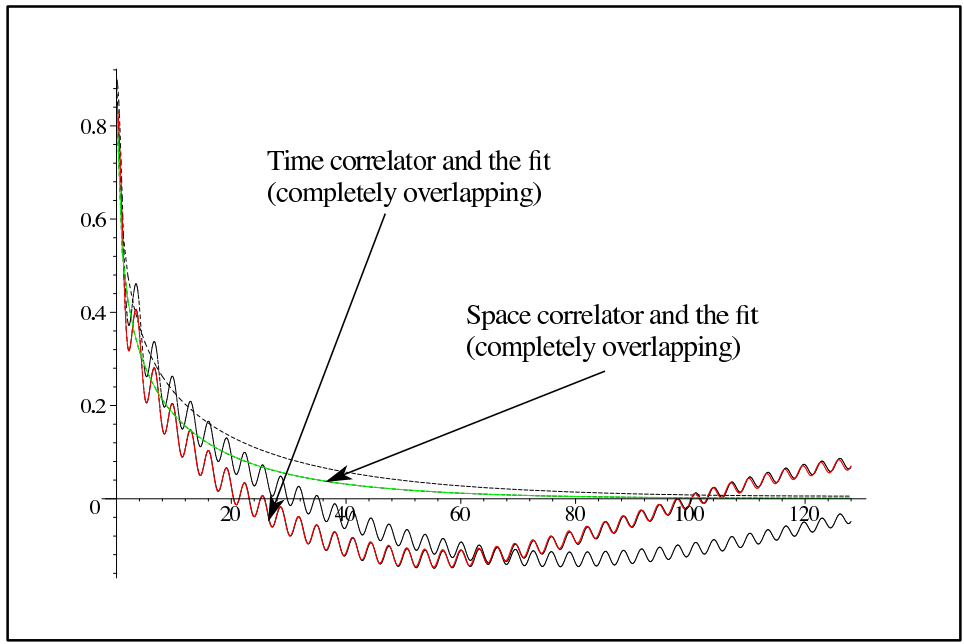}
  \end{center}
  \vspace{-3ex}
 \caption{{\bf The time and space correlators versus position (or time) in lattice units.} The time (wavy) and space (smooth) correlators are shown for the same parameters as Fig.~(1). Again we show the free correlators for comparison.}
  \label{figb}
\end{figure}

For an example we use $N=256$, $m_0^2=51/N^2$, $\lambda=20/N^2$, and $t_f=3N/4$. We generate $5000$ trajectories with an initial mass parameter $\mu=m_0$,
and obtain the correlation functions at
$t_f$. A least squares fit to free correlators with an adjustable mass produces a more appropriate input mass $\mu$. We produce another 5000 trajectories with this $\mu$ as input, and then obtain a new $\mu$. This process is repeated until
the input mass agrees with the extracted mass. We find that it
converges fast, typically in three interations.   In this case the iteration converges to the value
$m^2=100.0/N^2$. All three correlators have excellent fits to the corresponding free correlators with this mass. This is shown in Figs.~(\ref{figa}-\ref{figb}). For comparison we have also displayed the significantly different free correlators with $\mu=m_0$, to clearly show the effect of the mass renormalization.

We shall find that almost all of this mass shift, to within a percent or two, is due to the one-loop renormalization. We will have to go beyond one-loop in the following. However this example has already demonstrated some important features of the classical simulation.
\begin{itemize}
\item The system tends to evolve towards the renormalized physical mass from whatever mass is used to specify the initial condition. 
\item The masses from the various correlators agree. This is an indication that Lorentz symmetry breaking due to lattice effects is not significant.
\item The physical mass can be determined to high accuracy.
\end{itemize}

\section{A Precision Measurement of Mass}

The one-loop graph calculated on a lattice with discrete space and continuous time, after a Wick rotation, is given by
\begin{eqnarray}
\Pi(m) & = & \frac{1}{N}\sum_{k}\int \frac{dk'}{2\pi }\frac{1}{k'^{2}+4\sin^2(\pi k/N)+m^{2}}\nonumber\\
 & = & \frac{1}{2N}\sum_{k}\frac{1}{\sqrt{4\sin^2(\pi k/N)+m^2}}.\label{onel}\end{eqnarray}
This one-loop mass renormalization effect can be used to define the one-loop gap equation,
\begin{equation}
 m^2_{\mathrm{gap}} = m^2_0 +3\lambda \Pi(m_{\mathrm{gap}}).
\end{equation}
By obtaining the self-consistent mass $m_{\mathrm{gap}}$ from this equation we effectively sum
up graphs with chains of bubbles, where each bubble represents $\Pi$. From the
gap mass we may define a dimensionless coupling $g\equiv\lambda/m_{\mathrm{gap}}^{2}$, and the extent to which the gap equation can be trusted will be determined by this coupling. The gap equation provides a useful representation of the
physics for quite a large region of the $m_0^{2}$-$\lambda$
plane. This is because the one-loop graph is the only (log) divergent graph in this $1+1$ dimensional theory, and it thus dominates.

For the previous example we have $g=0.2$ and $m_{\mathrm{gap}}^2=101.7/N^2$, to be compared with the mass extracted from the simulation of $m^2=100/N^2$.\footnote{Since $m_{\mathrm{gap}}$ is much closer to $m$ than is $m_0$, in practice we use $m_{\mathrm{gap}}$ as the initial trial mass $\mu$ to start the simulation.} Given that there are two-loop corrections to the gap mass that are not included in the one-loop gap equation, our goal then is to decide whether the difference between our extracted mass and the gap mass is consistent with the size of the two-loop corrections from quantum field theory. We will first need to discuss the errors and uncertainties inherent in the classical simulation, and work to reduce them, in order to obtain the required accuracy. Then we need to determine what the quantum corrections actually are, and for this we will consider both lattice Monte Carlo methods and lattice perturbation theory.

At least at weak coupling we have seen that the evolving classical field configuration quickly settles down to a stable configuration that incorporates the effects of interactions. The stability of an interacting system with a nonthermal energy spectrum (an $\hbar\omega$ spectrum) is itself nontrivial, but at least in the continuum the $\hbar\omega$ spectrum is at a Lorentz invariant symmetry point, and thus stability may be a consequence. But on a lattice there is Lorentz symmetry breaking induced by the lattice itself, and this is expected to lead to thermalization. Thermalization of classical fields on a lattice is a well studied phenomena \cite{thermstable}. And consistent with these studies, our results indicate that the thermalization time scale is very long compared to the time scales of interest to us; on these shorter time scales, and for weak to moderate coupling, the interacting quantum-like configurations are very nearly free of thermalization effects. Presumably the lattice Lorentz symmetry breaking does not strongly affect the physics below the lattice cutoff.

Given the observed stability at weak coupling we can ask how far we should evolve forward in time to obtain the correlators. We are motivated to average over a fairly long range of time for the following reason. When we extract the mass for different choices of $t_f$ we find that the extracted mass fluctuates slightly as a function of $t_f$. For the example above it fluctuates within a $\pm 0.5\%$ band. This indicates that the configuration is not completely stationary, even if it is stable. But since the mass fluctuates around a stable central value, we may average the masses obtained over a range of $t_f$ and thus obtain an accurate mass.

There is more of a problem for large couplings. Here we find that the central value can slowly drift in one direction as $t_f$ increases to large values, in which case we are unable to obtain a reliable mass.  This could be taken as a sign of mild thermalization effects.\footnote{At couplings significantly larger than one these effects are more pronounced, and the amplitudes of the lowest modes, and in particular the zero-mode, can be seen to grow with $t_f$.} It is also sensitive to the size of the mass; for masses, $m^2N^2\gtrsim80$, the drift for increasing $t_f$ is towards higher mass, while for $m^2N^2\lesssim80$ it is towards lower mass.\footnote{This is for intermediate size coupling; for couplings significantly larger than one the drift is towards larger mass.} For $m^2N^2$ up to a few hundred the drift doesn't show up until $g\gtrsim 0.6$

We have also seen that we can extract mass from three different correlators. Of the three we find that the space correlator is the least reliable. It shows the most sensitivity to numerical errors and $t_f$ dependence. This is related to the fact that it is an exponentially decreasing function, and thus a fit is sensitive to only a small spatial range inversely proportional to the mass. The zero-mode correlator $G_0(i)$ and the time correlator $G_t(i)$ tend to agree well with each other. But the pure sinusoidal oscillations exhibited by $G_0(i)$ for small couplings tend to become damped oscillations at large coupling, due to the mild thermalization effects.

We thus concentrate on the time correlator $G_t(i)$.
Our prescription is to average over twenty equally spaced values of $t_f$ up to a final time $t_f=10 N$ (a temporal extent 10 times the spatial extent) to obtain a mass. We then average between $10N$ and $20 N$ to obtain a second mass. If the second mass does not deviate significantly from the first, then we accept the first mass as a reliable extracted mass.  We also note that the time correlator contains useful information in its deviation in functional form from the best fit free correlator, this effect becoming visible at larger coupling (it is not visible in the weak coupling correlators in Figs.~(\ref{figa}-\ref{figb})). These deviations in form are expected since the mass corrections at higher loops are momentum dependent.

\section{Comparison to Quantum Methods}

We now turn to the extraction of the mass in a lattice quantum theory. We would like to use an action and a lattice as similar as possible to the classical theory. However, standard lattice methods for the quantum theory make it necessary to Wick rotate
the time direction, and a periodic boundary condition is
imposed along it as well, giving a periodic Euclidean space.
Correspondingly we take $a_t=a=1$ rather than the $10a_t=a=1$ in the classical simulation. Then the
action for lattice quantum field theory is
\begin{eqnarray}
 S &=& \sum_{i, j = 0}^{N-1} 
  \frac{(\phi(i+1,j)-\phi(i,j))^2}{2}
  +\frac{(\phi(i,j+1)-\phi(i,j))^2}{2}
  +\frac{m_0^2}{2}\phi^2(i,j) +\frac{\lambda}{4}\phi^4(i,j) .
\end{eqnarray}

The different lattice regularization causes an obvious difference
in the value of the one-loop self-energy graph (to be compared with (\ref{onel})), which now becomes
\begin{eqnarray}
 \Pi_\mathrm{E}(m) = \frac{1}{N^2} \sum_{k, k'=-N/2}^{N/2-1}
  \frac{1}{4\sin^2(\pi k/N)+4\sin^2(\pi k'/N)+m^2}.
\end{eqnarray}
The gap equation for this theory (where the subscript E stands for the Euclidean lattice) is
\begin{equation}
 m^2_{\mathrm{Egap}} = m^2_0 +3\lambda \Pi_\mathrm{E}(m_{\mathrm{Egap}}).
\end{equation}
The resulting values $m_{\mathrm{Egap}}$ and $g_{\mathrm{E}}$
will thus differ from $m_{\mathrm{gap}}$ and $g$ from our previous gap equation.

\begin{figure}
\begin{center}\includegraphics[%
  scale=0.6]{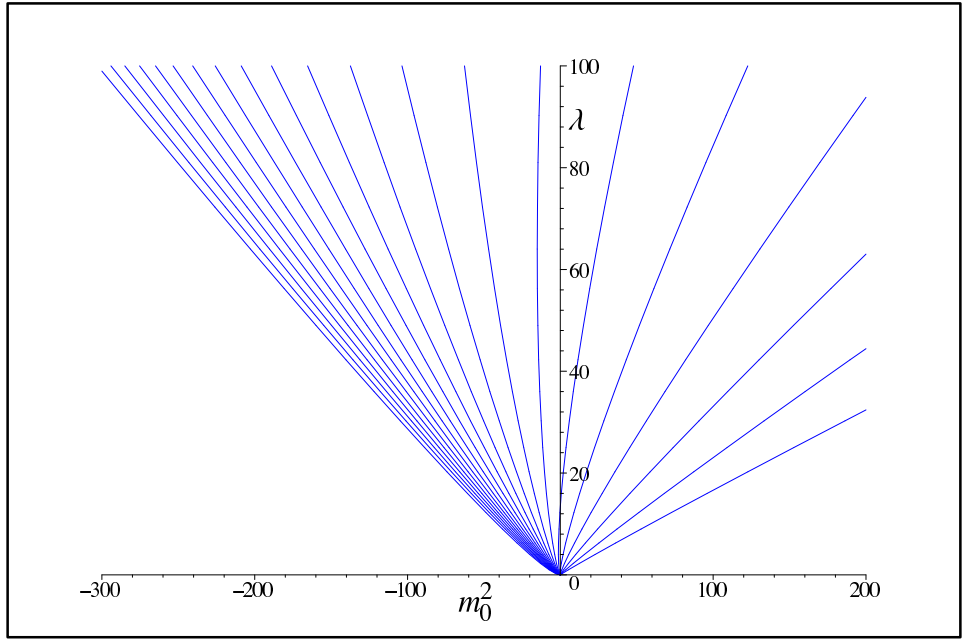}\end{center}  \vspace{-3ex}
\caption{\label{F}
Lines of constant $g$, with $\lambda$ and $m_0^{2}$
in units of $1/N^{2}$ ($N=256$). The $g$ of adjacent lines
decreasing from left to right by a factor of 0.8. The largest coupling
is the critical coupling of $g=10.24$ and the smallest is $g=0.12$.}
\end{figure}

\begin{figure}
\begin{center}\includegraphics[%
  scale=0.6]{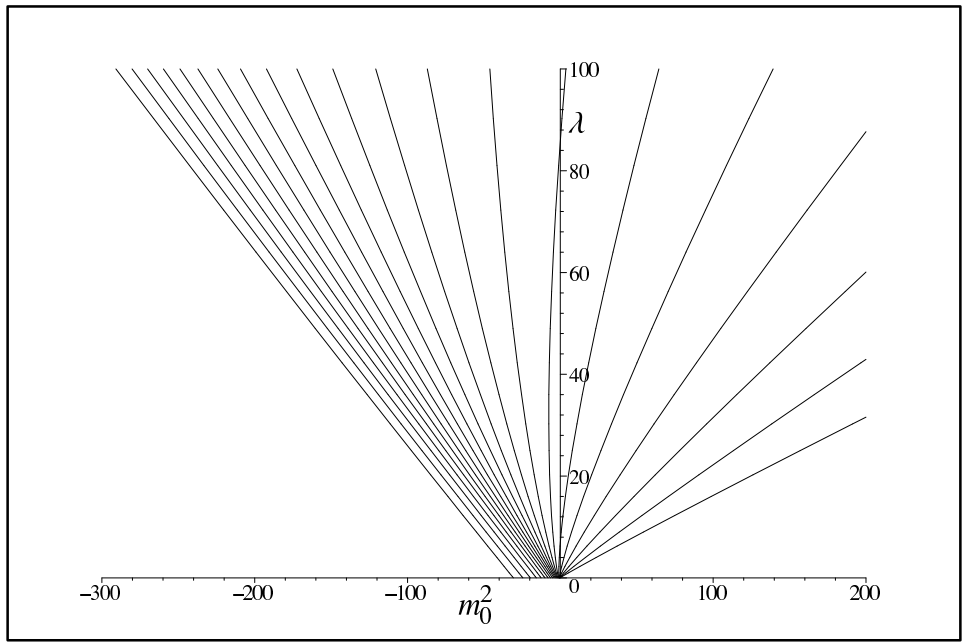}\end{center}  \vspace{-3ex}
\caption{\label{Fb}The same as Fig.~(\ref{F}), but for $g_{\mathrm{E}}$.}
\end{figure}

It is useful to graphically compare the results of these gap equations. Fig.~(\ref{F}) and (\ref{Fb}) show the behavior of lines of constant $g$ and $g_{\mathrm{E}}$ in the $m_0^2$-$\lambda$ plane. We see that large $g$ occurs when $m_0^{2}$ is sufficiently
negative. The largest coupling line on this graph is $g=10.24$,
which is the value of the critical coupling as determined by previous
Monte Carlo simulations of the quantum theory \cite{Lnz97}.\footnote{The critical coupling obtained from density matrix renormalization group methods \cite{Sug04} is 9.98.} In contrast to the
gap mass, the physical mass is expected to vanish on this line. We see that the $g_{\mathrm{E}}=10.24$
line has quite a different behavior at small $\lambda$ since it does
not intersect the origin. The reason for the difference is that $\Pi_{\mathrm{E}}(m)\rightarrow1/m^{2}$ as $m\rightarrow0$ while $\Pi(m)\rightarrow1/m$.

\begin{figure}
\begin{center}\includegraphics[%
  scale=0.6]{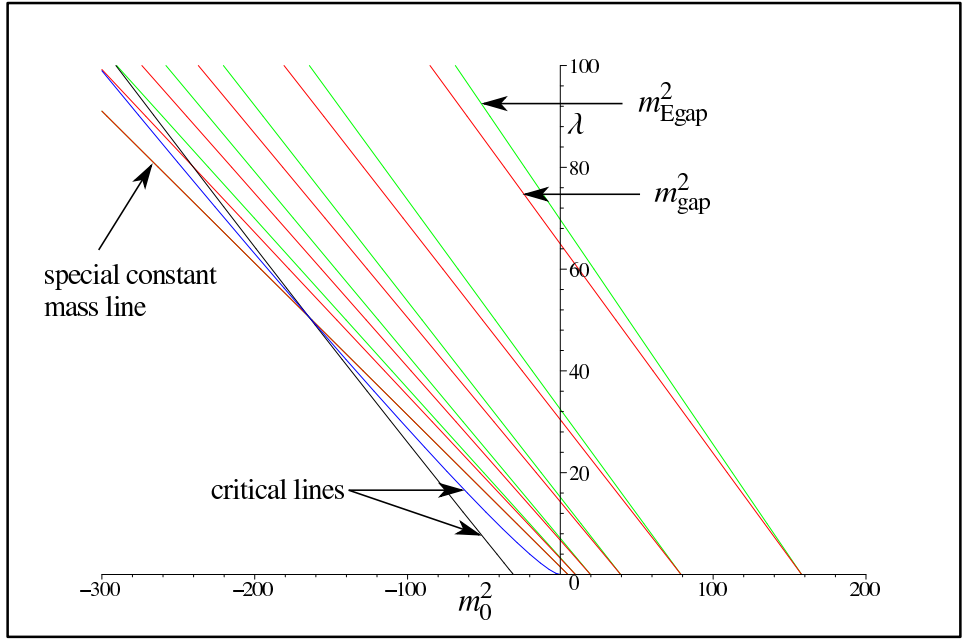}\end{center}  \vspace{-3ex}
\caption{\label{G}Lines of constant $m_{\mathrm{gap}}^{2}$ and $m_{\mathrm{Egap}}^{2}$ in units of $1/N^{2}$. The mass values are given by the intersection
of these lines with the $x$-axis. Also shown is a special constant
mass line and the critical coupling lines of the two theories.}
\end{figure}

We further illustrate the differences in the two lattice regularizations in Fig.~(\ref{G}),
where we compare the lines of constant gap masses. We see that the
differences basically grow linearly with $\lambda$. We would like to account for this
obvious source of difference when extracting the physical masses.
We therefore concentrate on the deviation of the physical mass
in each theory from the respective gap mass. The leading order deviation
from the gap mass is a finite two-loop effect, and
we expect that the lattice dependence of this leading order
difference will be small. This gives us reason to focus our study of
the physical masses at weaker coupling.

Also displayed in Fig.~(\ref{G}) is a special line of constant gap
mass, where the gap masses of the two theories agree. For $N=256$
this occurs at $m_{\mathrm{gap}}^{2}=m_{\mathrm{Egap}}^{2}=4.9338/N^{2}$. And since the gap masses agree, $g=g_\mathrm{E}$ along this line. Thus in the vicinity of this line the obvious differences between the two theories are minimized. In particular the two theories would agree on the location of the critical point along this line.

In our lattice quantum simulations we start with a hot start, and as our
updating scheme we choose the heat-bath algorithm followed by a Wolff
step to reduce critical slowing down. `One update' of the lattice consists of visiting each site with
the heat-bath algorithm and then applying a Wolff single-cluster
step \cite{Wlf89}. We monitor the relative size of the generated Wolff clusters, and we ensure that the system has thermalized by monitoring the action. The autocorrelation time $\tau$ is estimated by measuring the autocorrelation function $\rho(t)$ of various operators, such as $\phi(0,0)\phi(0,N/2)$ and the action, and computing the integrated autocorrelation time according to $\tau_{int}=\frac{1}{2}+\sum_{t=1}^{M}\rho(t)$. $\rho(0)=1$ and $M$ is such that for $t>M$, $\rho(t)$ is essentially noise.

 The physical mass is obtained by fitting
the correlation function
\begin{eqnarray}
 G_E(i) &\equiv& \left\langle \frac{1}{N^3}\sum_{i', j', k'=0}^{N-1}
  \phi(i',j')\phi(i+i',k')\right\rangle
\end{eqnarray}
to a $\cosh$ or a decaying exponential. Here $\langle\cdot\rangle$
is the thermal expectation value over different configurations. The systematic errors are the most difficult to control. To monitor these we compare results at neighboring values for parameters where similar results are expected. We also repeat the same simulations with different random number generator seeds.

As a check on the quantum simulation we find that it does give physical masses compatible with the gap equation, to the extent that the gap equation can be trusted. The mass extracted at strong coupling is also consistent with the expected critical behavior. The decrease of the physical mass towards zero as $g_\mathrm{E}$ is increased serves as one method for the determination of the location of the critical line. For another, the order parameter \[\Phi=\frac{1}{N^2}\sum_{i, j=0}^{N-1}\phi(i,j)\]
and its histogram can be studied as the size of the lattice is changed (see \cite{Lnz97}). The same data can be used to extract a critical exponent $\nu$ defined by
\begin{eqnarray}
 m_{phys} \propto |\lambda -\lambda_c|^{\nu}.
\end{eqnarray}
In all these studies we obtain results from our quantum simulation consistent with previous studies.

But the quantum simulation proved to be rather deficient at producing the type of precision that is needed in our study. A meaningful comparison to the classical theory requires that we accurately determine the difference between the physical mass and the gap mass. For a $256\times 256$ lattice, and for a run time of the order of a few days to extract a mass at a single ($m_0$, $\lambda$) point, a large percentage error still arose in the mass difference.\footnote{Through the use of a multi-processor cluster we have looked at roughly 50 points, using 50000 configurations for each.} This error seems to be related to problems with correlated data and/or autocorrelation, and it is exacerbated for smaller physical mass. This ruled out any meaningful comparison of the theories along the special line $m_{\mathrm{gap}}^{2}=m_{\mathrm{Egap}}^{2}=4.9338/N^{2}$ that we identified above. But as we shall see shortly, the quantum simulation results will still serve a useful purpose.

\section{The Two-Loop Effect}

We turn to improving the gap equation itself. Although the gap equation is nonperturbative in the sense that it resums an infinite
set of graphs, it omits some effects of second order and higher in
the coupling. At leading perturbative order the difference between the physical mass and the gap mass is determined by a single two-loop diagram, the ``sunset'' diagram:\\\begin{center}\includegraphics[scale=.9]{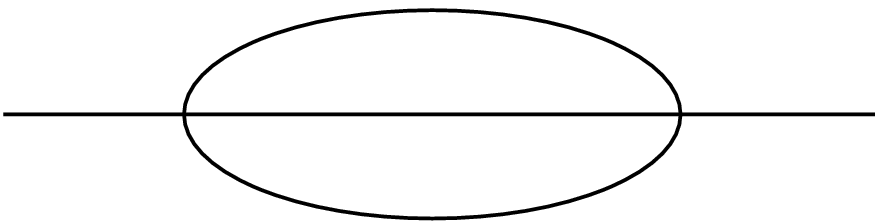}\end{center} In the lattice version of
the quantum theory, on the square Euclidean lattice, this diagram is given by an explicit multiple summation,
\begin{eqnarray}
\Pi_{2\mathrm{E}}(m,k) & = & \frac{1}{N^{4}}\sum_{\{p_{1},p_2,q_1,q_2\}=-N/2}^{N/2-1}G(p_{1}+k,p_{2})G(q_{1},q_{2})G(p_{1}+q_{1},p_{2}+q_{2}),\\
G(p_{1},p_{2}) & = & \frac{1}{4\sin^2(\pi p_{1}/N)+4\sin^2(\pi p_{2}/N)+m^{2}}.\end{eqnarray}
Here $k$ represents an external momentum. This calculation gives a correction to the 2-point function at what corresponds
to space-like external momentum, since this is a Euclidean quantum theory, and we find that $\Pi_{2\mathrm{E}}(m,k)$
monotonically increases as $k\rightarrow0$. The classical theory however has Lorentzian signature, and so we actually need $\Pi_{2}$ at the time-like on-shell momentum. We will use $\Pi_{2\mathrm{E}}(m,0)$ as an estimate for this.\footnote{By naive extrapolation from space-like momenta it is an underestimate, but probably by less than 10\%.} Moreover our calculation is on the square Euclidean lattice, rather than the lattice of the classical simulation, but this additional error  should be small.

For a $g^2$ improved gap equation to be useful we need to have some idea of the range of $g$ over which it is accurate, before even higher order effects become important. This is where the quantum lattice results are useful. Although there are large errors on the extracted mass correction at any single point in parameter space, when taken collectively the quantum simulation data supports $\Pi_{2\mathrm{E}}(m,0)$ as an estimate of the difference between the physical and gap masses. This appears to be a good estimate for couplings as large as $g\approx 0.6$. As a further check we have directly estimated the single three-loop graph using the same technique as above (for smaller lattices) and found that it is $\lesssim g/2$ times the size of the two-loop graph.

We therefore consider a two-loop gap equation of the following form,
\begin{equation}
 m^2_{\mathrm{gap}} = m^2_0 +[1-\varepsilon(m_\mathrm{gap})]3\Pi(m_{\mathrm{gap}})\lambda-6\Pi_{2\mathrm{E}}(m_{\mathrm{gap}},0)\lambda^{2}.
\label{TL}\end{equation}
We have inserted a correction factor in the $\lambda$ term to account for possible lattice artifacts or finite size effects inherent in our classical simulation.\footnote{In particular there is the effect responsible for the slightly fluctuating nature of the mass as a function of $t_f$ that we have noted earlier.} It is important to account for this because a small correction in the $\lambda$ term can compete in size with the $\lambda^2$ term. Similarly a correction to the $\lambda^2$ term of similar size can safely be ignored. This gap equation can be considered as a quantum model to be tested against the classical simulation data.

Our strategy then is to use (\ref{TL}) as a one parameter model for the extracted physical mass $m_\mathrm{phys}$ from the classical simulation.  We run the simulation for a range of values of $m_0$ and $\lambda$ that produce values for $m_\mathrm{phys}$ very close to a fixed $m_\mathrm{gap}$, and thereby determine $\varepsilon(m_\mathrm{gap})$ through a best fit. We choose the couplings $g=(0.1,0.2,0.3,0.4,0.5)$. We use 5000 trajectories for each determination of a mass. For the four values $m_\mathrm{gap}^2=(60, 80, 100, 140)/N^2$ we find $\varepsilon(m_\mathrm{gap})\approx (0.032,0.029,0.029,0.029)$.

\begin{figure}
\begin{center}\includegraphics[scale=0.8]{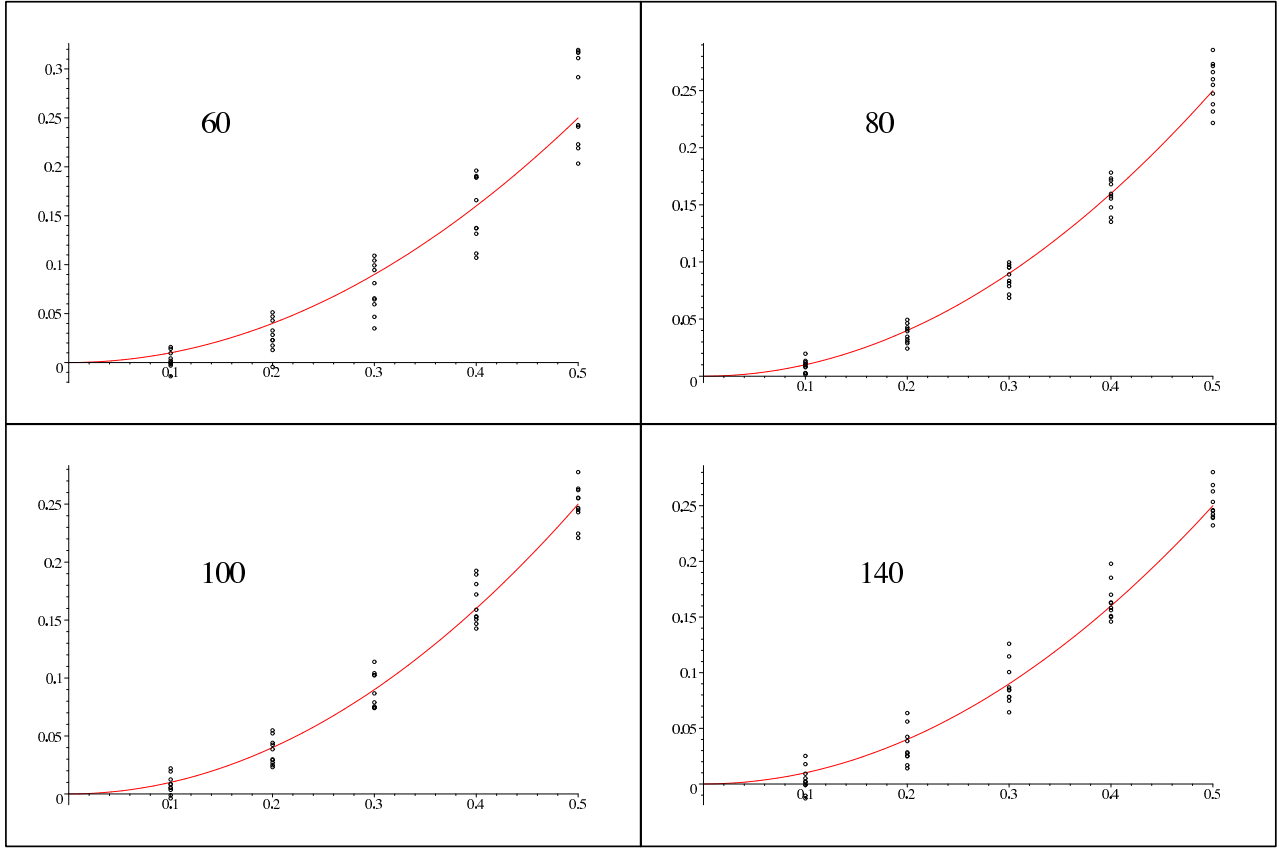}\end{center}  \vspace{-3ex}
\caption{\label{fige} {\bf Plots of the r.h.s of Equation (\ref{gg}) versus $g$ showing a comparison between the classical simulation and the quantum model.  } The graphs are labelled by $m^2N^2$ and the solid line is the $g^2$ quantum field theory prediction. Each plot displays the results of 10 different random number seeds.}
\end{figure}

To test the quantum model for $m_\mathrm{phys}$ we write it in terms of the dimensionless coupling $g=\lambda/m_\mathrm{gap}^2$ and then isolate the $g^2$ dependence to get
\begin{equation}
g^2 = \frac{m^2_{\mathrm{phys}} -\{m^2_0+[1-\varepsilon(m_\mathrm{gap})]3\Pi(m_{\mathrm{gap}})m^2_{\mathrm{gap}}g\}}{-6\Pi_{2\mathrm{E}}(m_{\mathrm{gap}},0) m^4_{\mathrm{gap}}}.
\label{gg}\end{equation}
We then determine the r.h.s from the classical simulation for a fixed $m_\mathrm{gap}$ and for a range of $m_0$ and $g$ that give $m_\mathrm{phys}\approx m_\mathrm{gap}$. This is displayed in the plots in Fig.~(\ref{fige}) where we see the residual quadratic dependence on $g$, and compare to the superimposed $g^2$ line of the quantum model. (The points would lie along a straight line if there was no $g^2$ effect.)

The spread of the points for 10 different random number seeds gives an indication of the errors. Of the four choices of $m_\mathrm{gap}$, the errors in the $m_\mathrm{gap}^2=80/N^2\;(60/N^2)$ plot are the smallest (largest), and for larger masses the errors again grow larger. In section 3 we have already noted that $m_\mathrm{gap}^2=80/N^2$ suffers least from the mass drift phenomenon at strong coupling, and thus the pattern of errors observed here are correlated with that phenomenon. In fact the drift is found to occur for $g\gtrsim 0.4$ in the $m_\mathrm{gap}^2=60/N^2$ case,\footnote{According to our previous prescription in section 3 the $g=0.4$ and 0.5 points should not be considered reliable.} and we can see how the sign of the drift tends to skew the $g^2$ effect towards a larger value.

The other three plots show that the results of the classical simulation are very  well described by the quantum model, where the coefficient of the $\lambda^2$ term is given by the sunset diagram of quantum field theory. This is our main result. We can also consider a two parameter fit to the data in these three plots, where we allow the coefficients of both the $\lambda$ and $\lambda^2$ terms to vary. This yields a $\lambda^2$ coefficient $\approx 8\%$ larger than the quantum prediction. For an independent set of data for masses $m_\mathrm{gap}^2 N^2$ from 60 to 200, where we only considered values of $g$ up to $0.3$, we obtained values for the $\lambda^2$ coefficient that deviated between $-2\%$ and $13\%$ from the quantum value. All these deviations are similar to the theoretical error in our estimate of the two-loop graph on mass shell.

\begin{figure}
\begin{center}\includegraphics[%
  scale=0.6]{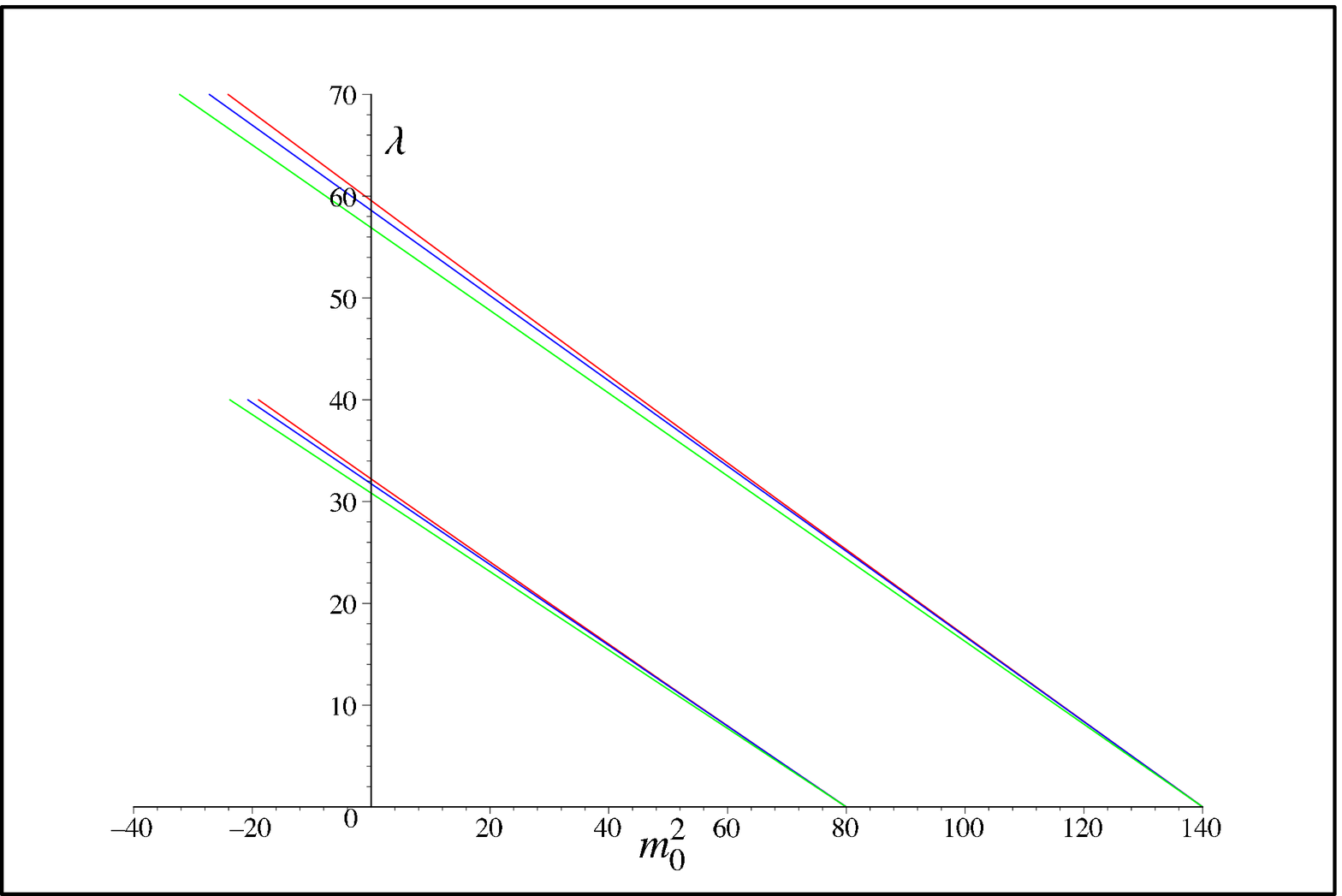}\end{center}  \vspace{-3ex}
\caption{\label{figc} Lines of constant mass, for $g$ up to 0.5. The bottom line in each case comes from the original gap equation, the next includes the $\varepsilon$ correction, and the top includes the $\lambda^2$ correction as well.}
\end{figure}

It is instructive to consider again lines of constant $m_{\mathrm{gap}}$ on the $m_0^2$-$\lambda$ plane, where we now use the corrected two-loop gap equation (\ref{TL}). In Fig.~(\ref{figc}) we compare the result to the original one-loop gap equation, by showing the effect of the $\varepsilon$ correction with no $\lambda^2$ term, and then the effect with the $\lambda^2$ term added as well. This figure highlights the fact that the simulation has to be highly accurate to see the small $\lambda^2$ effect. We have attained this level of precision in the classical simulation, and we have come nowhere close to it using conventional Monte Carlo methods in the quantum theory.

\section{Summary}

We have discussed the extraction of quantum loop effects from a classical simulation. We considered the time evolution of a scalar field configuration realizing the zero-point energy
spectrum of quantum field theory and including an averaging over phases. The time evolution incorporates the effects of interactions and the trajectory quickly approaches a stable configuration on the time scale of interest, at weak to moderate
couplings. The physical masses obtained from either the time or space correlation functions are in agreement, as would be expected by Lorentz invariance. Our method is clearly distinct from quantum
simulations and uses only a tiny fraction of CPU time. This fraction is roughly $10^{-4}$.\footnote{We compare times to produce comparable accuracy, 500 trajectories in the classical simulation versus 50000 configurations in the quantum simulation.}

In this way we have been able to perform a precision measurement of the effects of two-loop corrections within a classical simulation, with results that are consistent with quantum field theory. In \cite{Bob05} two of us studied the classical simulation at strong coupling and looked at two signatures of critical behavior, the overlap of the space and time correlators signaling masslessness and the development of a vacuum expectation value for the field. The findings were consistent with the location of the critical line in quantum field theory.

On the other hand in \cite{Bob05} it was also indicated that the loop expansion of the classical theory was not identical to the quantum theory; and it was suggested that the sunset diagram, the particular two-loop diagram of interest here, may have missing contributions in the classical theory. Due to uncertainties our results here may still be compatible with missing contributions, but they imply that the missing contributions in this case are accidentally small or vanishing. More generally, any perturbative discrepancy is not easy to reconcile with the success of the classical simulation at both strong and weak coupling. However this puzzle is resolved, we still find the concept of a classical simulation recovering most or all of a two-loop effect to be rather surprising and unexpected.

Note added: For further work on resolving this puzzle see \cite{H}.

\section*{Acknowledgments}
We thank Joel Giedt for helpful discussions and suggestions, and we are grateful to Randy Lewis for making available computational resources in Regina. This work was supported in part
by the National Science and Engineering Research Council of Canada.

\end{document}